\documentclass[
aps,%
12pt,%
final,%
notitlepage,%
oneside,%
onecolumn,%
nobibnotes,%
nofootinbib,%
superscriptaddress,%
noshowpacs,%
centertags]%
{revtex4}
\begin{document}

\title{ALICE FIT data processing and performance during LHC Run 3}

\author{\textcopyright 2020 A. Maevskaya for the ALICE Collaboration}
\email{alla@inr.ru}
\affiliation{Institute for Nuclear Research, RAS, Moscow, Russia}

\begin{abstract}
During the upcoming Run 3 and Run 4 at the LHC the
upgraded ALICE (A Large Ion Collider Experiment) will operate at a
significantly higher luminosity and will collect two orders of magnitude
more events than in Run 1 and Run 2. A part of the ALICE upgrade is the
new Fast Interaction Trigger (FIT). This thoroughly redesigned detector
combines, in one system, the functionality of the four forward detectors
used by ALICE during the LHC Run 2: T0, V0, FMD and AD. The FIT will
monitor luminosity and background, provide feedback to the LHC, and
generate minimum bias, vertex and centrality triggers, in real time.
During the offline analysis FIT data will be used to extract the precise
collision time needed for time-of-flight (TOF) particle identification.
During the heavy-ion collisions, FIT will also determine multiplicity,
centrality, and event plane. The FIT electronics designed to function
both in the continuous and the triggered readout mode. In these
proceedings the FIT simulation, software, and raw data processing are
briefly described. However, the main focus is on the detector
performance, trigger efficiencies, collision time, and centrality
resolution.
\end{abstract}

\maketitle

\section{INTRODUCTION}

ALICE (A Large Ion Collider Experiment) is the dedicated heavy ion
experiment at the CERN Large Hadron Collider (LHC)~\cite{Collaboration_2008}. The main
goals of ALICE are: physics of the strongly interacting matter at
extreme energy densities and the formation of the Quark-Gluon Plasma
(QGP) --- a new phase of matter. The ALICE detector is undergoing a major
upgrade during the Long Shutdown 2 (2019-2021). The main reason for the
upgrade is the increased luminosity and interaction rate. The LHC will
deliver Pb--Pb collisions at up to luminosity
$6 \cdot 10^{27}cm^{-2}s^{-1}$,
corresponding to an interaction rate of 50 kHz. The goal of ALICE is to
integrate a luminosity of 13~$nb^{-1}$ for Pb--Pb
collisions at $\sqrt{s_{NN}}$ = 5.5 TeV, together with dedicated
p--Pb and pp reference runs. Data from pp collisions will also be
collected at the nominal LHC energy $\sqrt{s}$ = 14 TeV~\cite{Abelev_et_al_2014}. Run 3 at the
CERN LHC is scheduled to start in 2022.

The ALICE upgrade includes:

\begin{itemize}
\item
  A new, high-resolution, low material Inner Tracking System (ITS)
\item
  An upgrade of the Time Projection Chamber (TPC)
\item
  New Muon Forward Tracker (MFT)
\item
  New Central Trigger Processor (CTP)
\item
  New Fast Interaction Trigger (FIT) detector
\item
  New readout and trigger systems allowing for continuous data taking.
\end{itemize}

\section{FIT DETECTOR}

\subsection{General description}

\begin{figure}
\setcaptionmargin{5mm}
\onelinecaptionstrue
\includegraphics[width=\textwidth]{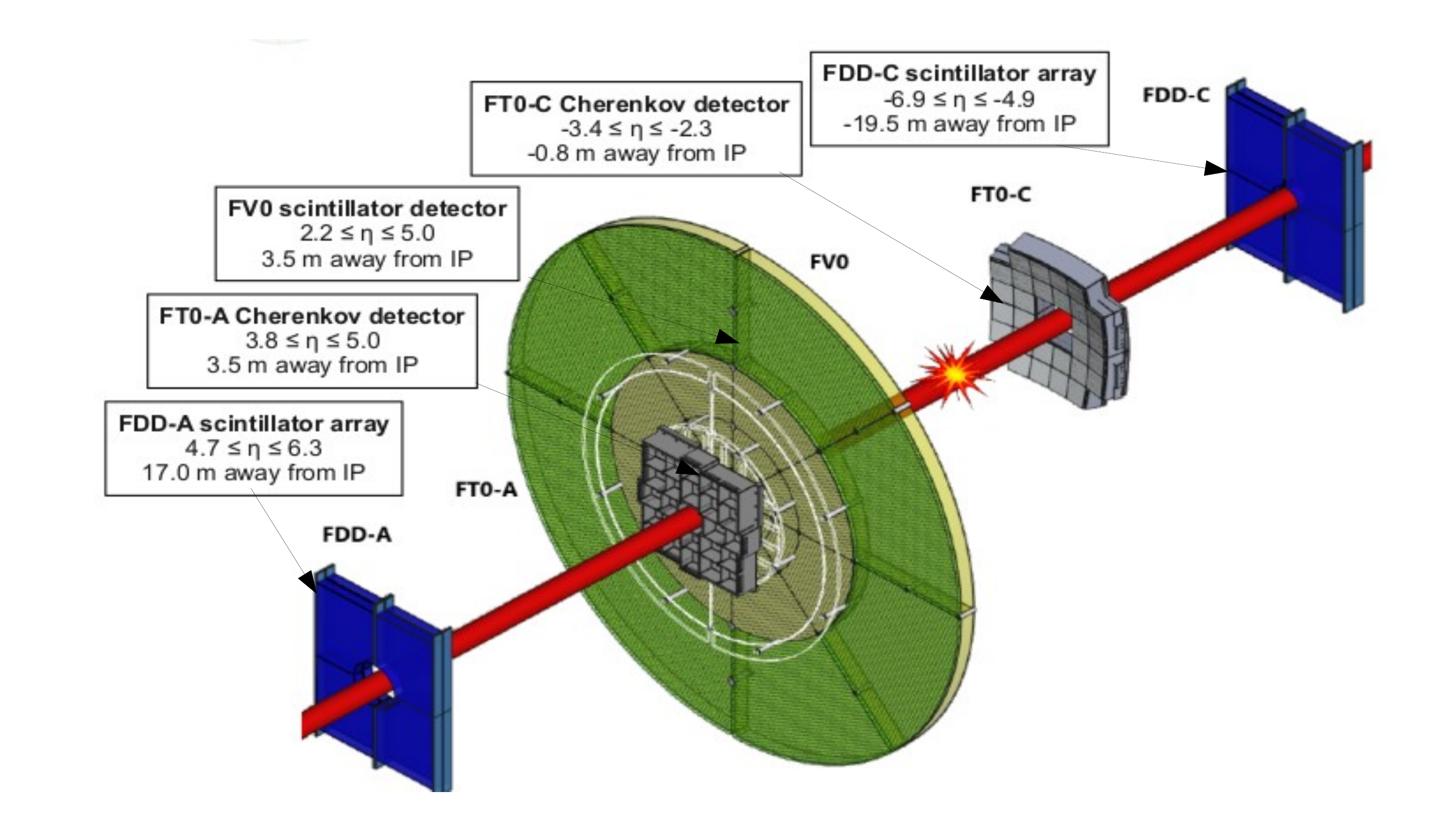}
\captionstyle{normal}
\caption{FIT detector layout.}\label{fig:layout}
\end{figure}

The new Fast Interaction Trigger~\cite{Cortese:781854} replaces four Run 2 detectors
(T0, VZERO, FMD and AD~\cite{Trzaska_2017}) with three new subdetectors: FV0, FT0
and FDD. These three subdetectors, each utilizing a different
technology, are placed at both sides of the interaction point, in the
forward and backward rapidity regions, as shown in Fig.~\ref{fig:layout}. The online
functionality of FIT includes luminosity monitoring and the
lowest-latency (<~425~ns) minimum bias, vertex, and centrality
triggers. Offline, FIT provides the precise collision time for the
TOF-based particle identification, determines the centrality and event
plane, and measures the cross section of diffractive processes. In
addition, FIT can reject beam-gas events and provide vetoes for
ultra-peripheral collisions.

\subsection{Construction of FIT subdetectors}

The FT0 subdetector has two modular arrays, FT0A and FT0C, placed at
opposite sides of the interaction point. The FT0A array consists of 24
modules, and the FT0C array, of 28 modules. Each module has four,
optically separated, 2~cm-thick quartz Cherenkov radiators, coupled to a
customized PLANACON MCP-PMT. The anodes of the MCP-PMT are grouped into
four outputs providing an independent readout channel from each
individual radiator segment. As a result, FT0A delivers 96 readout
channels and FT0C, 112 channels. The FT0C, being located close to the
interaction point, has a concave shape to equalize the flightpath of
primary particles and assure their perpendicular entry to the radiators.
The intrinsic time resolution of each section of a module is
$\approx 13$~ps. The FT0 contributes to the minimum bias trigger,
luminosity monitoring, and background rejection.

The FV0 is a large scintillator disk, assembled from 40 optically
insulated elements, arranged in 8 sectors and 5 rings with progressing
radii. Clear optical fibers deliver light from each element to a
Hamamatsu R5924-70 PMT. Owing to its much larger size, each sector of
the outermost ring is read by two PMTs. In total there are 48 FV0
readout channels. Time resolution is $\approx 200$~ps. The FV0
provides inputs for minimum bias and multiplicity triggers at LM (Level
Minus one) level and, because of its large acceptance, delivers data for
centrality and event plane determination.

The FDD consists of two stations, FDDA and FDDC, covering very forward
rapidity region at the opposite sides of interaction point. The stations
are made of two layers of plastic scintillators, divided into four
quadrants. Each quadrant has two wavelengths shifting (WLS) bars
connected to individual PMTs via a bundle of clear optical fibers.
Benefiting from the forward pseudorapidity coverage, the FDD will
contribute to cross section measurements of diffractive processes~\cite{Abelev_2013} and studies of ultra-peripheral collisions, and participate in
beam monitoring and beam-gas rejection.

\subsection{FIT electronics}

The FT0, FV0, and FDD utilize the same electronics scheme based on two
custom-designed modules: the processing Module (PM) and the Trigger and
Clock Module (TCM). The PM processes and digitizes input signals, packs
the data for readout (in continuous or triggered mode), and makes the
first stage calculations for trigger decision. The TCM processes data
from PMs, makes the final trigger decisions, provides accurate clock
reference, and serves as the slow control interface to the connected
PMs.

\section{FIT DATA PROCESSING}

All ALICE detectors are integrated into a common Detector Control System
and Online and Offline Computing system called
$\text{O}^2$~\cite{Buncic:2011297}.

The functional flow of the $\text{O}^{2}$ system includes a
succession of steps. Data arrive at the First Level Processors (FLP)
from the detectors. The first data compression is performed inside of an
FPGA-based readout card (Common Readout Unit). The data is transferred
from the detectors either in a triggered or in a continuous mode.
Temporary simulated raw data were used for system preparation and
performance studies. Heart-Beat triggers from Central Trigger Processor
(CTP) are used to chop data in Sub-Time Frames (STF). The STF are
assembled into Time Frames (TF) in the Event Processing Nodes (EPN). One
TF packet includes 128 or 256 orbits. A second step of data aggregation
is performed to assemble the data from all detector inputs. A global
calibration, the first reconstruction and data compression using
Graphics Processing Unit (GPU) is performed synchronously with the data
taking. Compressed Time Frames (CTF) are stored permanently on tapes.
Results of each step are monitored within the Quality Control (QC)
framework.

In the asynchronous stage, a second (and possibly third) reconstruction
with final calibration is run on the $\text{O}^2$ EPN farm and
on the GRID. The final Analysis Object Data (AOD) is produced and stored
permanently.

\section{FIT FT0 PERFORMANCE}

\subsection{FIT FT0 performance in pp collisions at $\sqrt{s}$ =14 TeV}

The FT0 can produce trigger signals every 25 ns that is for each LHC
bunch crossing. However, due to the limited acceptance, the efficiency
has to be verified by simulations. Pythia8~\cite{Sj_strand_2008} was used to simulate
particles from pp collisions. Twenty thousand pp collisions were
generated and transported through the ALICE setup. Cherenkov photons
from relativistic charged particles traversing quartz radiators were
utilized to produce digitized signals taking into account the detector
response and possible pile-up. The signals were used to evaluate the
efficiency of the following FT0 triggers:

FT0A -- signal only from the A side;

FT0C -- signal only from the C;

Vertex -- signals from both sides and vertex within given range.

\begin{figure}
\setcaptionmargin{5mm}
\onelinecaptionstrue
\includegraphics[width=\textwidth]{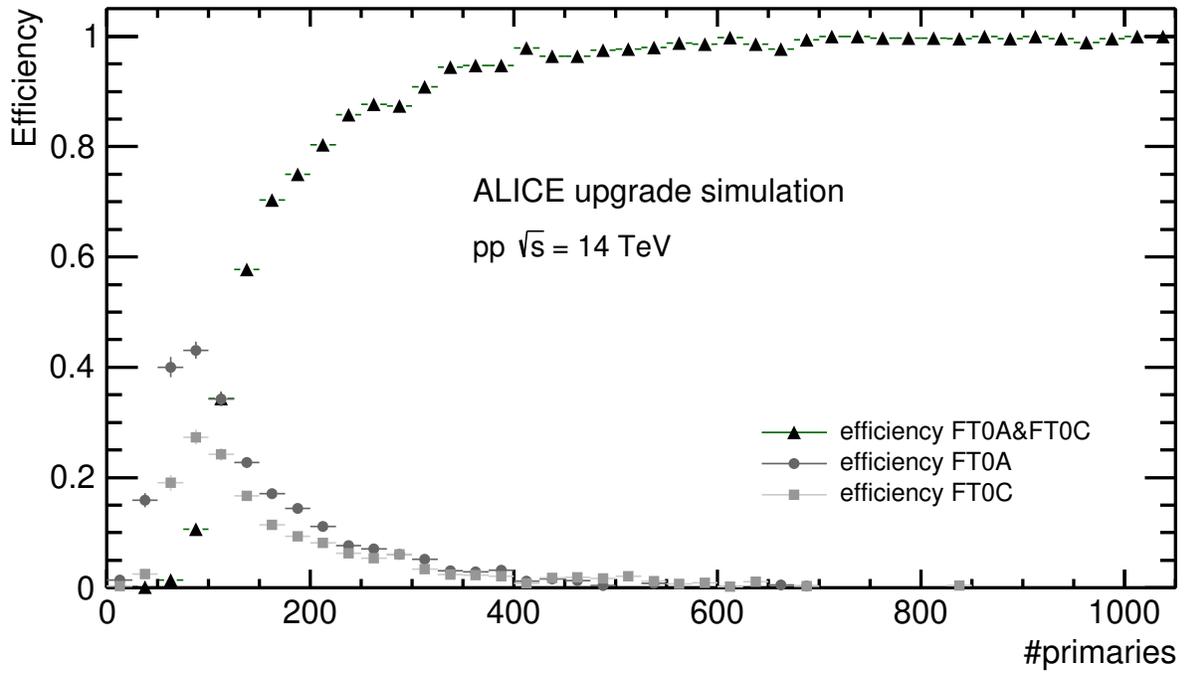}
\captionstyle{normal}
\caption{Efficiency of FT0 triggers: vertex trigger (triangles), trigger only
  from A side (circles); trigger only from C side (squares).} \label{fig:eff_pp}
\end{figure}

Fig.~\ref{fig:eff_pp} shows FT0 trigger efficiencies as a function of event
multiplicity. Total vertex trigger efficiency is approximately 77\%,
fraction of triggered events with only FT0A trigger signal is 9\%, FT0C
-- 6\%.

\begin{figure}
\setcaptionmargin{5mm}
\onelinecaptionstrue
\includegraphics[width=\textwidth]{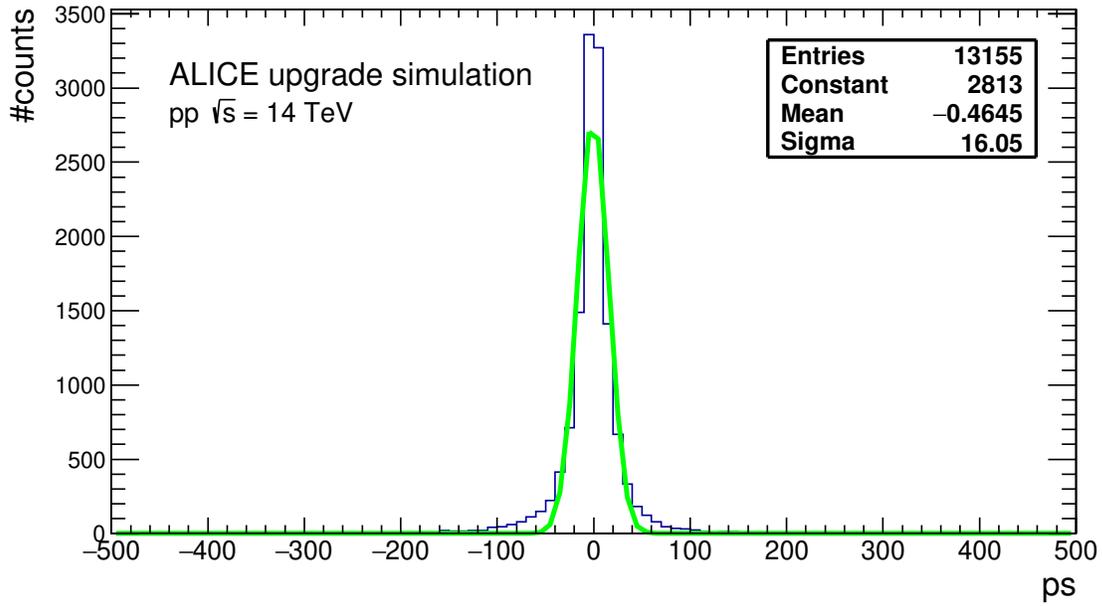}
\captionstyle{normal}
\caption{Distribution of a difference between average arrival
times on each side corrected with the primary vertex for pp collisions
at $\sqrt{s}$ =14 TeV. The bold line is a Gaussian fit, with the sigma of the
fit being the resolution on the collision time.}\label{fig:resolution}
\end{figure}

Collision time is half of the sum of the average arrival times at FT0A
and FT0C and does not depend on the position of the primary vertex. The
resolution can be estimated as a sigma parameter of a Gaussian fit of
the distribution of differences between the average arrival times on
each side, corrected with the primary vertex. As shown in Fig.~\ref{fig:resolution}, the
resolution of the collision time is below 20~ps for pp collisions at $\sqrt{s}$~
=~14 TeV. If the collision time measured can not be obtained due to the
limited acceptance of FT0, the individual times measured by FT0A and
FT0C, corrected for the primary vertex position, could be used.

\subsection{FT0 performance for Pb-Pb collisions}

When the impact parameter of two colliding ions is larger than the sum
their radii, hadron collisions are replaced by electromagnetic
interactions corresponding to photon-photon and photon-nuclear
collisions. The main source of background comes from pair production
($e^+e^-$), having orders of magnitude
larger cross section than the hadronic processes. For instance,
according to PYTHIA8, the Pb-Pb hadronic cross-section at
$\sqrt{s_{NN}}$~=~5.5~TeV is 8~b while the cross section for
electromagnetic collisions, used by the QED generator, developed
especially for ALICE, is around 180~kb. Fortunately, QED events have a
very low charged particle multiplicity. They can be rejected by setting
a threshold value for the sum of the FT0A and FT0C amplitudes.

\begin{figure}
\setcaptionmargin{5mm}
\onelinecaptionstrue
\includegraphics[width=\textwidth]{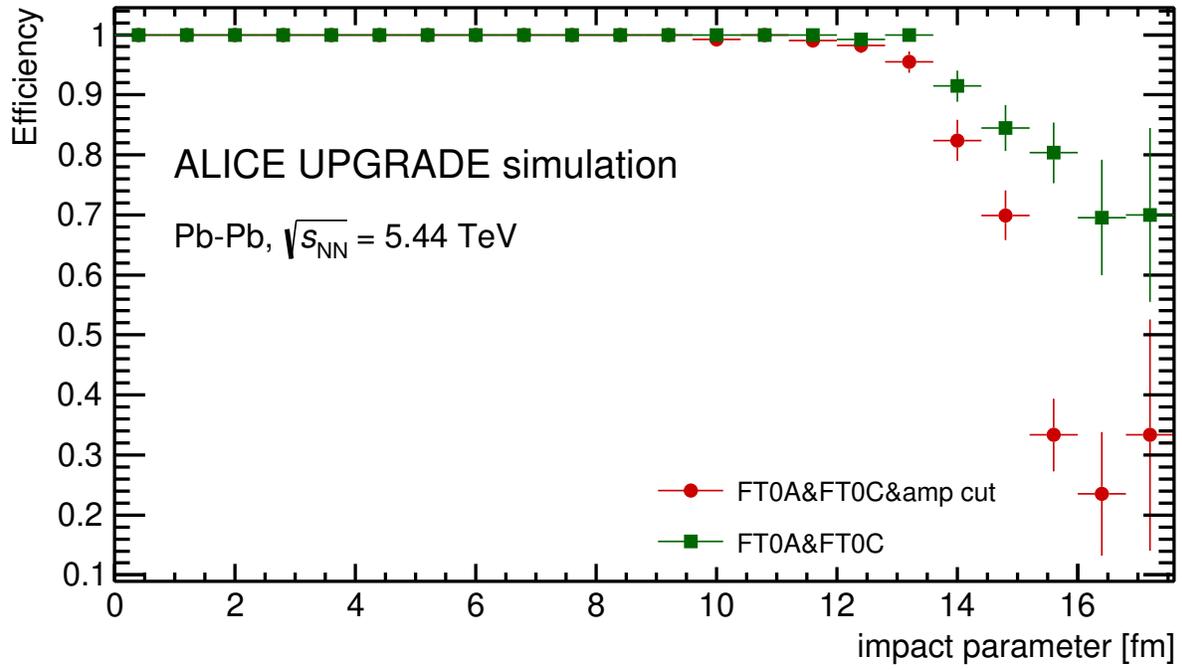}
\captionstyle{normal}
\caption{Minimum bias trigger efficiency as a function of impact
parameter. Red squires show efficiency of minimum bias trigger, green
circles are efficiency of both minimum bias trigger and cut of the sum
of the amplitude at 30 ADC channels}\label{fig:eff_pb}
\end{figure}

Fig.~\ref{fig:eff_pb} shows the efficiency of the minimum bias trigger (coincidence
between FT0A and FT0C) as a function of the impact parameter for hadron
collisions. The squares (dots) show the results with (without) the
selection on the amplitude. It is clear that, for events with an impact
parameter below 12~fm, the amplitude cut does not affect the efficiency
of the minimum bias trigger. The total efficiency of the FT0A and FT0C
trigger is $\approx 92\%$. The vertex trigger (coincidence
between FT0A and FT0C together with the requirement for the position of
the z-vertex lie to within 10~cm around interaction point) efficiency is
83\%. For central and semi-central events the efficiency of the vertex
trigger is 100\%.

\begin{figure}
\setcaptionmargin{5mm}
\onelinecaptionstrue
\includegraphics[width=\textwidth]{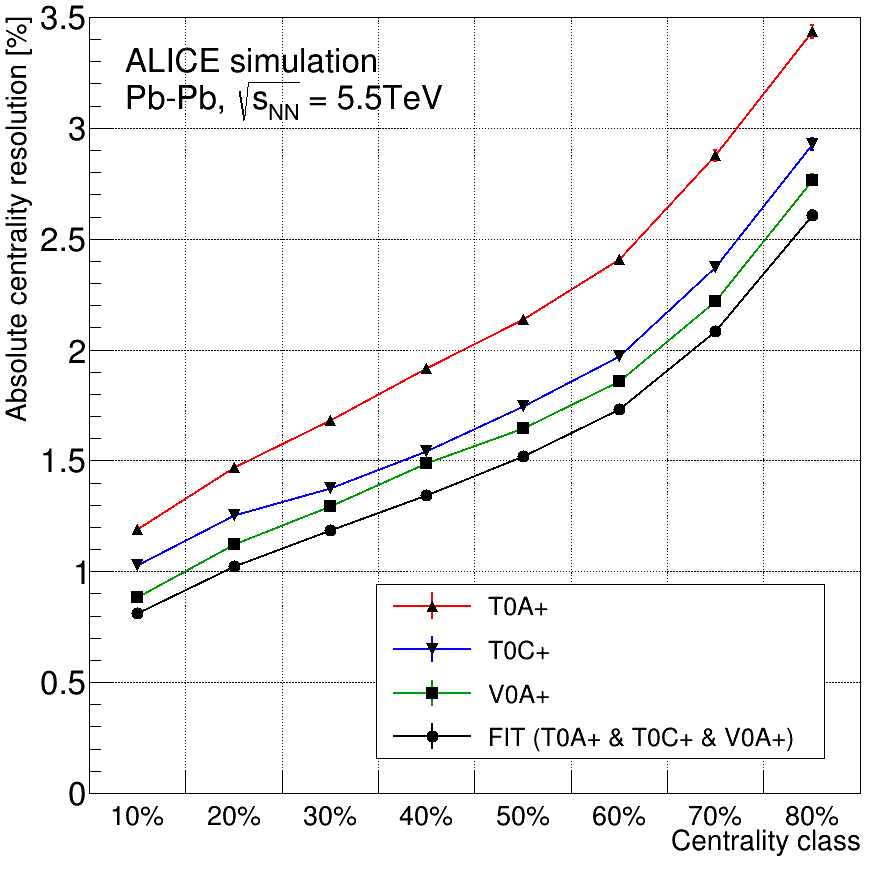}
\captionstyle{normal}
\caption{Centrality resolution for measurements with different
parts of the FIT detector. FT0A (up triangles), FT0C (down triangles) ,
FV0 (squares) , combination of all 3 (dots) .}\label{fig:centrality}
\end{figure}

Centrality determination with a good resolution is an important
functionality of the FIT detector. Fig.~\ref{fig:centrality} shows the centrality
resolution for Pb-Pb collision at $\sqrt{s_{NN}}$~=~5.5~TeV
calculated for FT0A, FT0C, and FV0 separately, and the combined
resolution (FT0A+FT0C+FV0).

\section{CONCLUSIONS}

Our analysis has demonstrated that the simulated performance of the FT0
subdetector of FIT satisfies the design requirements of the ALICE
experiment:
\begin{itemize}
\item The minimum bias trigger efficiency matches that of the VZERO detector
operated during the Run~1 and 2 of the LHC;
\item The collision time resolution is better than that of the T0 during the
Run~1 and Run~2;
\item The vertex trigger has a 100\% efficiency for semi-central and central
events.
\end{itemize}

\bibliography{Maevskaya_article}

\end{document}